# Autonomy by Design: Preserving Human Autonomy in AI Decision-Support

Stefan Buijsman[1,*]
Sarah E. Carter[2]
Juan Pablo Bermúdez[3,4]

[1] Ethics and Philosophy of Technology Section, TU Delft, Delft, The Netherlands
[2] Web Information Systems, TU Delft, Delft, The Netherlands
[3] Department of Philosophy, University of Southampton, Southampton, UK
[4] Facultad de Ciencias Sociales y Humanas, Universidad Externado de Colombia, Bogotá, Colombia

[*] Corresponding author. Email: S.N.R.Buijsman@tudelft.nl

**Abstract:** AI systems increasingly support human decision-making across domains of professional, skill-based, and personal activity. While previous work has examined how AI might affect human autonomy globally, the effects of AI on domain-specific autonomy—the capacity for self-governed action within defined realms of skill or expertise—remain understudied. We analyze how AI decision-support systems affect two key components of domain-specific autonomy: skilled competence (the ability to make informed judgments within one's domain) and authentic value-formation (the capacity to form genuine domain-relevant values and preferences). By engaging with prior investigations and analyzing empirical cases across medical, financial, and educational domains, we demonstrate how the absence of reliable failure indicators and the potential for unconscious value shifts can erode domain-specific autonomy both immediately and over time. We then develop a constructive framework for autonomy-preserving AI support systems. We propose specific socio-technical design patterns—including careful role specification, implementation of defeater mechanisms, and support for reflective practice—that can help maintain domain-specific autonomy while leveraging AI capabilities. This framework provides concrete guidance for developing AI systems that enhance rather than diminish human agency within specialized domains of action.

## *1. Introduction*

Artificial Intelligence systems are increasingly used to provide outputs that aim to facilitate decision-making, often by providing direct suggestions for a course of action. This decision support can happen discretely in the form of alerts, such as the EPIC system used by hospitals to detect risk of sepsis (a life-threatening reaction to infection) that are intended to trigger clinicians to take action (Habib et al., 2021). It can happen continuously, as with self-driving cars that do (most of) the work of driving the vehicle for you, but with the expectation that the human stays attentive to potential automation failures. And it can happen over time, the way Large Language Models (LLMs) are used to write initial drafts of texts then reviewed by people, and then again to finetune subsequent versions. All these interactions are meant to empower people in specific action domains. They should lead to better patient care, safer driving, and better-written outputs.

However, as we will argue, such uses of AI can affect human autonomy: the ability to make decisions and act according to values that are significantly 'one's own' (Christman, 2020; Prunkl, 2022). While there are more expansive ways of understanding the concept,[1] there is widespread agreement that autonomy involves some level of *self-governance*, which has two main conditions. On the one hand, *competency conditions* involve having the psychological capacities underlying practical rationality like memory, imagination, planning, self-control, and emotional regulation. On the other hand, *authenticity conditions* require that the motivations and values that drive the agent are truly their own, as opposed to distorted or illegitimately influenced from outside.

Researchers have examined how AI use can impact different aspects of people's autonomy (Burr et al., 2018; Prunkl, 2024), as well as our psychological need for feeling autonomous (Calvo et al., 2020). These analyses tend to focus on the impact of AI-supported decisions on those affected by them (like job candidates affected by hiring algorithms (Vaassen, 2022) or incarcerated people affected by automated parole decisions) (Rubel et al., 2021; E. Taylor, 2024). When examining the decision-makers themselves, researchers have focused on AI's impact on its impact on *global autonomy*: the overall level of autonomy a person enjoys across multiple spheres of action. Some have focused on AI's impacts on specific domains of autonomy, like interaction with social robots (Formosa, 2021). Here we focus on a particularly relevant and under-explored domain of human action: *skillful decision-making*, and the use of AI systems as tools for decision support.

We will argue that both competence and authenticity can be affected in AI decision-support interactions, which we understand as interactions in which the human uses AI outputs as input for a decision-making process. This ranges from medical doctors making diagnosis or prescription decisions and judges making parole decisions, to professionals deciding what feedback to provide to a colleague and drivers making motor decisions about how to steer a vehicle. Recent work has already explored the effects of algorithmic decision support on the autonomy of *those affected by* automated decisions; for example, how algorithmic management affects worker autonomy (Unruh et al., 2022); or how recommender systems impact autonomous value formation (Kasirzadeh & Evans, 2023). Instead, this paper focuses on the impact of algorithmic decision support on the autonomy of *the decision-makers* assisted by these technologies.

[1] Particularly, researchers have proposed that 'autonomy' might be a multi-dimensional concept, where not all of its dimensions are exhausted by self-governance. (See particularly (Killmister, 2018; Mackenzie, 2014) for more expansive multi-dimensional accounts.) In this paper we focus on self-governance as a shared core element of autonomy across different perspectives, and do not mean to claim that this exhausts the discussion: there is more to be said about how human-AI interactions affects other dimensions of autonomy.

We argue that decision-maker autonomy is at risk along both dimensions of autonomy, within the domains in which the AI provides support. Regarding competency, we highlight AI opacity as an issue for domain-specific competency during the use of AI (section 2.1) and deskilling as an issue for domain-specific competence in the long run (section 2.2). And regarding authenticity, we argue that interactions with AI systems to achieve one's goals can bias value-formation processes (section 2.3) leading to a change of values within the domain that is inauthentic. Importantly, all these effects can be domain-specific: the decision maker's autonomy can be affected *in the AI-supported decision domain* while their autonomy remains intact in other domains. This also means that not all losses of autonomy are problematic: if a coder gives up part of their autonomous ability to write code due to long-term interactions with AI coding assistants, then that may be unproblematic if their ability to produce code together with AI assistance is improved compared to the original autonomous ability. In many cases, dependence on technology reduces our autonomy in one task to improve our overall autonomy in achieving higher-level goals. However, many of the cases we discuss here relate to losses of autonomy in decision-making and control over our values, which is (we believe) much more problematic than a loss of autonomy in coding or mental arithmetic.

In part 3 we propose that these risks can be mitigated by intentionally designing the socio-technical environment within which AI decision support occurs. Supporting domain-specific competency requires providing sufficient information, particularly about risks of system failure, to counteract their opacity (section 3.1); appropriately distributing roles between humans and AI systems (section 3.2); designing training regimens in ways that protect skills (section 3.3). Additionally, the authenticity of domain-specific user values can be bolstered by choice frictions of the right kind (section 3.4). We thus formulate a number of socio-technical recommendations for the design of autonomy-preserving AI decision-support systems.

## *2. Challenges to Domain-Specific Autonomy in Human-AI Interactions*

### 2.1. Challenges to Synchronic Competency: The Lack of Failure Mode Warnings

AI systems, and especially deep neural networks, are widely considered to be epistemically opaque (Boge, 2022). We currently do not know why they produce the outputs that they do and therefore have no access to the reasons for these outputs. While we do of course have access to the AI systems themselves, the combination of a vast number of parameters with a lack of explicit representations (parameters typically cannot be linked to real-world quantities) means that when using AI systems we are often limited to outputs and accuracy scores. Research on Explainable AI aims to improve this situation (Ali et al., 2023), but as we discuss in section 3.1 still has a long way to go before it manages to provide effective explanations. However, users of AI systems lack more than just explanations of outputs. They also miss a broader set of indicators that tell us when an AI output is reliable and when it is not. In other words, we often cannot easily distinguish between cases when the AI we are using operates as it should, and cases when it may fail, requiring a shift to a more reflective engagement with— and perhaps an override of—the AI system's recommended path of action.

The absence of failure mode indicators affects how we relate to AI systems, in contrast with many other pieces of technology. From a technological mediation perspective (Verbeek, 2006), when we use a technology as a tool, this technology mediates both the way we act on the world and the way the world presents itself to us. As long as our use of the technology works as expected and is unimpeded, it is (in the terminology of Heidegger (Heidegger, 1996)) "ready-to-hand": the technology itself recedes from our attention and our engagement with it is intuitive and fluid. The flow of action is impeded only when we notice the tool is malfunctioning. This breakdown makes the technology appear as "present-at-hand": the immediate relationship between us and the world is interrupted, and the tool itself enters the focus of our attention. It is then that we can reflectively engage with it and critically assess why it is malfunctioning and what we should do about it. The skillful use of a tool requires that we appropriately manage the shift between intuitive, ready-to-hand engagements with the world and reflective, present-at-hand engagements with the tool itself. In AI systems, the lack of failure mode signals reduces the human decision-maker's ability to switch between intuitive and reflective modes, i.e., the ability to make the meta-decision of when to intuitively rely on the system's output and when to critically question it.

This kind of shift between intuitive and reflective modes is not as difficult for other technologies. For simple tools such as shovels we can immediately see when it does not function properly, as there will be visible damage to the tool even before we attempt to use it, and haptic feedback will highlight noticeable differences once we pick it up. While this certainly gets more complicated as tools get more sophisticated, even in the case of, e.g., an MRI machine there is a set of indicators and known failure modes for these machines that radiologists are trained to recognize. They can see, for example, when a patient is not holding their breath for the entire duration of a scan and will pause the scan as soon as (or even slightly before) this happens, as they know that this will lead to inaccurate imaging. Similarly, airplane automations come with long sets of instructions that pilots are trained on, complete with protocols for how to act when indicators light up, which pilots must know by heart. The system can of course still fail in unexpected ways, but in the prime cases where this has happened, such as the Boeing 737 Max incidents, that unexpected failure is recognized as a serious design error that should have been addressed both socially (in pilot training) and technically. Lack of clarity about failure modes fatally endangered the passengers of the airplanes and hindered the competency of the pilots flying the plane as they did not know what was failing in the plane or how to correct it.

AI systems are, in most setups, not accompanied by these kinds of warning signals, nor are their failure modes easily detected. There certainly are patterns to how AI systems fail. We know that they become less reliable for inputs that are dissimilar from the training data and we know that AI systems are sensitive to correlations in the data that are not always generalizable, the way object recognition systems often pay undue attention to color and texture, confusing other bright yellow objects with bananas (Hendrycks et al., 2021). Without support to spot dissimilar inputs, however, there is frequently no way for a user to tell when the system is operating outside its scope. Crucially, as we do not know what patterns the system is in fact sensitive to, there is no telling when it uses patterns that fail to generalize to the current situation.

One might hope here that we can indicate the uncertainty of an AI output and use that as a guideline for whether the system requires critical engagement. However, uncertainty quantification is still a serious challenge (Kuppers et al., 2020). Adversarial examples are a nice highlight of this fact, as they show that with minimal changes to the input we can get AI systems to produce wildly incorrect outputs with high confidence ratings (Alcorn et al., 2019).

This lack of failure transparency creates a situation where human decision-makers have to rely on general claims about the AI system's reliability. They have very limited tools to determine when to be critical about the functioning of the AI and when to accept it as is. This theoretical claim is reflected also in the empirical literature on decision support systems, where we see users struggle to calibrate their reliance on AI systems (Klingbeil et al., 2024; Schemmer et al., 2023). This can lead to either under-reliance on the system, as people decide that it is better to trust their own faculties, with lower performance as a result (He et al., 2023); or it can lead to overreliance, often accompanied by automation bias, where users adopt the AI's recommendation outputs even when they are incorrect (Klingbeil et al., 2024). Since AI systems are ideally instrumental to us (better) achieving our goals, this is a clear challenge to the competency dimension of autonomy.

Let us now go back to autonomy. This discussion of failure transparency highlights one aspect of the competency condition not so frequently discussed: *metacognition*, i.e., the set of abilities that allow an agent to monitor and control their own cognitive activities (Arango-Muñoz, 2011; Proust, 2019). By providing them with a sense of certainty or doubt, error or fluency, metacognition crucially enables decision-makers to reliably track to what extent their action advances their self-endorsed values and objectives within the chosen domain. If metacognition misfires, say, by providing us with a mistaken sense of certainty, or with a feeling of fluency dissociated from correct performance, we are no longer able to track whether we are making progress towards our goals in that domain — in short, without an acute sense of metacognition, our competency is significantly diminished. As a result, we under-rely or over-rely on the system making less competent decisions than if we were able to appropriately rely on the AI system.

Failure signals are thus crucial because by alerting us that (there is a risk that) something is wrong or out of the ordinary, they allow metacognitive feelings to adjust to the situation: a heightened sense of doubt and error allows us to steer information processing more cautiously. Without these signals, however, we tend to remain in an intuitive state of flow while making decisions that lead us astray from the values we intend to advance.

Thus, in human-AI interaction settings where a human decision-maker's access to reality is mediated by the output of an AI system that lacks failure signals, the decision-maker's autonomy is significantly compromised since they cannot reliably track when their decisions advance their values and when they do not. This reduction in competency is due to not having access to signals that allow them to metacognitively guide their cognitive processes.

### 2.2. Challenges to Diachronic Competency: Deskilling

Competency can also degrade over time, as we come to rely more and more upon AI systems to do parts of the work for us (OpenAI, 2023; Passi & Vorvoreanu, 2022). This kind of deskilling is not specific to AI: one of the best-known examples of deskilling is what the calculator has done to our mental math skills. However, with the wide range of tasks that we are considering to incorporate AI into, it is nevertheless important to be aware of the potential longer-term effects of this on our domain-specific autonomy. First, because there are recent suggestions that AI usage leads to a loss of critical thinking skills, harming the metacognition discussed above (Gerlich, 2025). Second, we also see domain-specific challenges: on the one hand, in cases where people are expected to keep monitoring the quality of decisions over time and, on the other hand, in cases where AI is expected to scaffold our skills in some environments, while we also have to make decisions in environments without AI support.

For the former, we can look at a range of cases where support through automation has lowered people's knowledge and their confidence in their own decisions. In a recent case, Wessel (Wessel, 2023) investigated the use of decision-support systems by financial professionals. These professionals reported that as reliance on AI increases, both their declarative knowledge of customer and financial data and their procedural knowledge of doing tasks without AI support decreased. Sutton et al. (Sutton et al., 2018) give a good overview of this effect in a wider range of domains. They mention that doctors have decreased confidence in their ability to diagnose patients after relying on AI systems (Goddard et al., 2011, 2014), that auditors with similar levels of experience are less able to perform audit tasks when they primarily worked with audit support systems (Dowling & Leech, 2014) and that marketing managers supported by machine learning show decreased levels of creativity (Wortmann et al., 2016). Together, these results suggest that sustained AI assistance can reduce the human decision-maker's domain-specific competency through time by altering both their *cognitive abilities* (making them less capable of acquiring and using decision-relevant information by themselves) and their *metacognitive processes* (reducing their sense of confidence in their ability to make appropriate decisions independently of the AI system's support). If compounded through time, these two effects can lead human decision-makers to over-rely on AI support, and make them less confident in rejecting the AI's suggestions when they should be overridden (OpenAI, 2023; Passi & Vorvoreanu, 2022). If you add up the AI system's lack of failure signals and the loss of human expertise, the result could be a significant overall loss in the capacity to track decision quality within expertise domains. This amounts to a loss in domain-specific competency over time, which would render humans less able to adequately evaluate the relevance of the AI's contributions, and thus make our decisions in those domains less autonomous.

Domain-specific deskilling takes place not only as the gradual loss of already-acquired skills: it also manifests in a reduced ability to acquire new ones. In a peer review task, the inclusion of an AI tool helped students give concrete, original and relevant suggestions. However, once the AI assistance was removed the participants did significantly worse than when assisted, providing shorter feedback that was flagged as needing revision more often, as well as writing more generic and unrelated comments. As the researchers themselves note: "AI prompts played a significant role in maintaining the quality of students' feedback" (Darvishi et al., 2024, p. 10). While perhaps not a case of deskilling per se, as the study did not look at long-term effects of using AI, it does highlight the risk that AI interaction may not support skill acquisition but rather generate dependence without actually acquiring an independent (autonomous) competence to provide high-quality feedback.

Deskilling through hindering new skill acquisition can also occur when engaging with AI chatbots, such as mental health chatbots. While the relationship between a mental health patient and their provider involves a level of dependence (Clemens, 2010), mental healthcare professionals have a duty to help someone who may struggle to advocate for themselves and recover their autonomy (Brown & Halpern, 2021). Mental health chatbots, however, do not replicate such a therapeutic relationship. As Hamdoun et al. (Hamdoun et al., 2023) write, these systems "present a paradox: the promise of 24/7 companionship and the expectation of self-sufficiency" (p. 32). While the claims of mental health chatbots to relieve acute psychological distress are widely supported in the literature, the relationship between mental health chatbot use and long-term psychological wellbeing is less clear (Li et al., 2023). In addition, studies with users of generic conversational agents have found evidence of users developing overdependent, sometimes arguably addictive, relationships to their chatbot (Haque & Rubya, 2023). For example, users of Replika, a companion conversational agent that also claims to improve mental health,[2] reported such dependence on their chatbots that they could no longer go for a walk without it or used it to replace real-world social interactions (Ma et al., 2024). These interactions with AI thus fail to provide users with the necessary social skills for maintaining psychological wellbeing.

In sum, the evidence discussed in this section suggests that AI decision support can lead human agents to lose competencies crucial to autonomous domain-specific choice through time because it can both erode previously-acquired decision skills and hinder the development of new ones. We highlight the twin risks of *cognitive deskilling*, where AI support prevents human agents from accessing the information required to maintain and develop skilled judgment, and *metacognitive deskilling*, where it leads human agents to lower their level of confidence in their AI-free decisions and judgments. These are all instances of a diachronic loss of the competencies necessary for domain-specific autonomous decision-making. Next, we look at the effects of AI support on the authenticity dimension of domain-specific autonomy before turning to design recommendations in section 3.

### 2.3. Challenges to Domain-Specific Authenticity

Traditionally, whether a person is authentic has been theorized as determined by whether their motivations and values are in the correct type of relation (e.g. whether an agent wholeheartedly endorses their motivations (Dworkin, 1988; Frankfurt, 1988)). Recently, Karlan (Karlan, 2024) has provided a seminal discussion about whether AI-supported decisions might lead human agents to decide inauthentically in that sense, i.e., going unwittingly against the values that they hold at the time of deciding . For example, to narrow 250 applicants for a junior job to a list of 5–6 interviewable ones, a hiring manager at a technology company uses an AI-based resume filtering system, and in doing so might have made a decision inconsistent with her commitment to being fair to all candidates, given the well-known proneness to exacerbate human biases of such systems (Deshpande et al., 2020; Gaddis, 2018).

While crucial, it has recently become clear that these 'static' accounts presuppose that the agent's values at the time of decision were formed authentically, and not themselves a product of inauthentic processes. Static accounts thus miss a crucial element of authenticity: our values and beliefs transform as our practical circumstances change and we develop our sense of who we are. Adopting a historical perspective (Christman, 2014; Mackenzie, 2014), a key question is whether the processes whereby an agent's values and beliefs change is authentic, or whether it alienates them from their identity. While there is not much agreement about what makes value shifts authentic (Feldman & Hazlett, 2013; C. Taylor,

[2] From https://replika.com/ : "Coaching: Build better habits and reduce anxiety," and a featured review from *The New York Times*: "Replika was designed to provide positive feedback to those who use it, in accordance with the therapeutic approach made famous by the American psychologist Carl Rogers, and many psychologists and therapists say the raw emotional support provided by such systems is real."

1992), one plausible basic condition is that the agent should be able to become aware of the value shift. That is because the person can judge to what extent the shift fits with who they are only if it is consciously accessible. More precisely, unless the agent is in principle able to become aware of a shift in their values, they cannot tell whether it supports or undermines their *practical identity* (i.e., the person's most central commitments and values which structure the way the agent sees the world, deliberates, and acts (Christman, 2014; Korsgaard, 1996)).

We extend the discussion of AI-aided decision making's impact on authenticity beyond its immediate effects (Karlan, 2024) and examine long-term value changes. In this section we will discuss the emerging evidence that AI decision support can generate shifts of values and beliefs of which the decision-makers remain unaware. We argue that if such inaccessible value shifts accumulate through time, the human decision-maker may change their value set in ways that erode autonomy by undermining authenticity diachronically, i.e. by leading the agent to adopt inauthentic values. In sum, because such cumulative value shifts are inherited in a *persistent* and *unconscious* manner, the decision-maker is unable to judge such shifts from the perspective of their practical identity to assess their authenticity.

To start, interacting with an AI system can result in shifts that *persist* even after the AI interaction has ceased. For example, Vicente & Matute (Vicente & Matute, 2023) found AI-inherited bias in a medical image classification task. Participants who had biased AI assistance made more errors (in line with the AI bias) on the classification tasks than those who had no AI assistance. In a second, unaided classification task, they found higher error rates in AI-assisted participants *even after* the AI system was removed. In addition, participants who reported greater levels of trust in the AI system made more errors than those with lower levels of trust.

While these studies only evaluated one unaided task after the initial AI interaction, we have reason to believe that such biases will persist for longer. This is because values, beliefs, and biases embedded in AI systems could be subconsciously transmitted to the user and influence their beliefs even after the system is no longer in use (Kidd & Birhane, 2023) since systems tend to offer biased information at a critical moment of curiosity, when someone is uncertain and open to learning, and after which it can be difficult to challenge. As Vicente and Matute found, overhyping the system's capabilities or presenting its results as fully objective further add to the likelihood that such a bias will be taken on by increasing trust.

Kidd and Birhane (Kidd & Birhane, 2023) further predict that, in such an open, trusting state of curiosity, it may not always be obvious to the person that such biases and beliefs are present. Emerging evidence suggests that such shifts are, indeed, *inherited unconsciously* by the decision-maker when interacting with an AI system. For example, Jakesch et al. (Jakesch et al., 2023) evaluated opinion posts about whether social media is good for society, where participants were assisted by an LLM biased for or against social media. Participants who used the writing assistant were more likely to write posts that reflected the assistant's bias and to mimic its bias when asked for their opinion on social media after the experiment when compared with a control group. Crucially, most participants were not aware that the model was skewed and could be influencing their writing. When asked, only 10% of participants detected the model's bias when it aligned with their existing opinion, and only 30% did when the bias contradicted it. In addition, most participants reported that they felt the assistant was knowledgeable and had expertise in the topic. In addition, interactions with social media recommendations can influence value and belief shifts unbeknownst to the user. Social media has become a widely-used source of information on relevant world events: many users interact with the recommender system to seek information relevant to them – for example, using social media posts to initially flag a relevant news event for the user to investigate in more detail elsewhere (Edgerly, 2017). However, it is well-accepted that recommender systems are often designed to retain user attention and engagement, the fuel of the "attention economy" (Davenport & Beck, 2001; Goldhaber, n.d.). Kasirzadeh and Evans (Kasirzadeh & Evans, 2023) found that systems readily promote polarized political content early in order to increase the acceptance of recommendations later on. This phenomenon of "user tampering" can arguably promote diachronic shifts in the user's values and beliefs for economic gains. Indeed, this change of interest in and acceptability of certain content could be viewed as a modification of a user's opinions to retain user engagement. Taken together, not only can such value shifts be persistent, but the human collaborator can unconsciously inherit them from the AI system.

Supported by this emerging evidence, we postulate that inauthentic value shifts within domains (e.g. political opinions, or bias in diagnosis) can occur through repeated interactions with AI systems. Over time, an AI system's existing biases and embedded value sets can be *unconsciously inherited by the decision-maker*, even when such a shift is in tension with the user's core values. Together, this threatens their ability to assess such shifts in relation to their practical identity. The lack of awareness that the shifts are occurring means that the decision-maker cannot actively endorse them, and this can result in problematic, cumulative shifts to the agent's beliefs, values, and attitudes over time.

Notably, those most at risk could be those marginalized from discussions around AI and its development. Many have flagged the prevalence of mostly WEIRD (Western, Educated, Industrial, Rich, Developed; (Henrich et al., 2010)) values in LLMs. For example, ChatGPT tends to align with American norms, values, and opinions (Benkler et al., 2023; Cao et al., 2023; Johnson et al., 2022). As such beliefs, as we have seen, could be plausibly unconsciously picked up by those interacting with AI systems, those of non-WEIRD backgrounds are especially at risk of acquiring values that are not in accordance with their practical identity. This is especially true as use of predominantly Western-orgin AI, including LLMs such as Chat-GPT, continues to rise in other parts of the world. Therefore, care must be taken to explicitly address the concerns of authenticity loss due to undetected value shifts. We will provide suggestions about this in section 3.3.

In this juncture, we think that it's relevant to briefly explore the link between our discussion and discussions of automated manipulation (Jongepier & Klenk, 2022; Susser et al., 2019). Manipulation comes up naturally when thinking of ways of undermining our autonomy. Different accounts of manipulation share the idea that the goal of manipulation is to get someone to take a certain action, while some put the emphasis on the manipulator's intentions (Susser et al., 2019) and others on the manipulator's carelessness (Klenk, 2022). Regardless of the specific account one adopts, we think that they link up naturally with our discussion of autonomy here. The stereotypical cases of manipulation, focused on individual actions, describe situations where we fail to act in accordance with our own values and reasons due to external factors. We are made to act differently than we would have, and our competence to act on our own values and reasons may be undermined. The examples above already illustrated this: we may end up with biases and opinions that we would, upon reflection, want to avoid. As a result, after sufficiently long histories of interaction with AI-based decision support systems, people taking actions due to their AI-inherited biases and beliefs that they otherwise would not have, and do so in predictable ways. We thus have an effective way to change behavior via the long-term erosion of authenticity.

This type of manipulation is, in our view, an especially worrying case. Manipulations that focus on singular actions, or that change our behavior through targeting the competency dimension (keeping our values intact but leading us to act differently than we would like to) are problematic but easier to fix. Competencies can be restored, perhaps sometimes even simply by removing the AI system from the process. Furthermore, barriers to competency are more easily detected (e.g. through introspection). On the other hand, once our values have been changed to something inauthentic, it is both harder to detect and much more difficult to repair. When manipulation of this kind has happened at a larger scale, how do we determine which values are actually authentic, and which are the result of manipulation? Moreover, regaining authenticity can face resistance from within, since the agent now has a competition between inconsistent sets of values. We therefore think that it is important to also pay close attention to the authenticity dimension of autonomy, including in the debate on manipulation. The recommendations in the next section are a starting point, but overall we believe that more attention is needed for diachronic manipulation that is harder to spot and affects us more deeply, yet is not captured as naturally by current accounts of manipulation that focus on whether a specific, singular, action is manipulated.

## *3. What Now? Recommendations*

We have argued that autonomy can be harmed in a variety of ways as a result of collaborative interactions with AI systems. Yet, we believe that these harms to autonomy can be mitigated. We need to consciously design the interactions we have with AI systems in order to protect autonomy, keeping up human competence and authenticity while having the benefits of AI support. In this section we look at what can be done to tackle the challenges of a lack of failure transparency, of deskilling and of opaque value shifts due to interactions with AI. While none of these can be tackled by changing the AI algorithm alone, a socio-technical lens at the broad setup within which we work with AI helps identify a way forward.

### 3.1. Protect Competency Synchronically by Making Failures more Transparent

The lack of failure transparency was introduced as stemming from the more general opacity of AI systems. Because we don't know why the system produces a certain output, it is very difficult to evaluate whether the output makes sense or to anticipate when the system as a whole is less reliable. As such, attempts to make AI systems more explainable (Ali et al., 2023) are a natural first step to tackling this challenge. However, while we believe that good explanations would be very valuable and despite a vast literature on the technical side of explainable AI (Guidotti et al., 2019), the current state of what XAI can offer users is insufficient. Empirical evaluations of popular explainable AI techniques have shown little to no cognitive benefits (van der Waa et al., 2021). Users do not make better decisions, i.e. do not disagree with wrong AI outputs or agree with correct AI outputs more often (and thus do not improve competence), based on this extra information (Labarta et al., 2024). They are also not able to better anticipate what the AI system will do for new inputs (Poursabzi-Sangdeh et al., 2021).

These empirical results warrant a cautious take on making AI explainable, at least for the currently used XAI techniques. Since the issue for competency stems from a challenge to metacognition, we would need to see explanations that support metacognition specifically. However, as the current techniques do not allow us to reason any better about AI outputs or the behavior of the AI system as a whole, we don't see this as a crucial support for metacognition: they do not (yet) provide the signals needed to reliably steer information processing and decision-making. While recent work is exploring new types of explainable AI techniques, using for example causal inference techniques (Beckers & Halpern, 2019; Buijsman, 2022; Millière & Buckner, 2024), these have yet to be tested empirically to see if they could support metacognitive competency in different domains. We therefore suggest to look further than explainable AI, although purely due to the current state of the art. If new, more effective, explanations are develeoped then explainability is still a promising way to improve competence, and if (as a reviewer suggested) they also highlight the values underlying the AI outputs then this can also help with authenticity.

More general approaches to tackling failure transparency that does not rely on making the AI system itself more transparent have also been suggested. Buijsman and Veluwenkamp (Buijsman & Veluwenkamp, 2022) discuss precisely this issue in light of *defeaters*: pieces of information that either give the user evidence to doubt the AI system's reliability (undermining) or offer evidence in favor of an alternative conclusion (undercutting). In both categories we can design the socio-technical system in such a way that users receive such defeaters/warning signals when appropriate. For example, users can receive a warning when an input is an outlier compared to the AI system's training set. Outlier detection methods are readily available (Boukerche et al., 2021) and we know that in general AI systems will be less reliable when processing outliers, thus giving grounds to provide the decision-maker with an undermining defeater. Users will then have extra reason to critically reflect on the AI system's output before incorporating it into their decision-making process. As a second example, there are bound to be reasons that make a system misfire. The NarxCare system, an AI system that predicts whether a patient requesting opioids might be addicted to them, screens people's entire medication purchase history (Pozzi, 2023). This can include painkillers bought for pets, with legitimate reasons that show that there is in fact no reason to suspect addiction, but the evidence that the painkillers were bought for a different reason is not processed by the system itself. Either a supporting system highlighting prescriptions from veterinarians or a sensitivity to patients explanations (and possible provision of harder evidence) would be a good way to build in an undercutting defeater in the process.

While empirical evidence is still wanting, the very purpose of defeaters is to highlight when a system might fail, either due to unreliability or due to insensitivity to additional reasons for a decision. We thus expect that designing the socio-technical systems that include these warning signals will aid people's competency in decision-making by supporting metacognitive accuracy, ultimately allowing decision-makers to distinguish cases when they should intuitively rely on AI outputs from those in which they should more critically evaluate them. If there is a warning signal/defeater then we know that we should switch our use of the AI system from the intuitive to the reflective mode. In addition, we believe that a broader redesign of socio-technical systems would be helpful as it is also often the role that is given to AI systems that leads to problems with competence, both through the lack of failure transparency and through deskilling.

### 3.2. Protect Competency Skills by Redesigning Socio-Technical Systems

A common setup of human-AI interactions is one where both the human and the AI are solving the same task. Decision support systems are a good example, as the AI system focuses on the same decision as the human. Self-driving cars are another example, as both the AI and the human have the task to drive the car. Yet in precisely these kinds of situations we see that it is hard for people to maintain their attention on the task at hand. Thus, for self-driving cars we see that the more the human trusts in the AI system, the longer their reaction times are (Payre et al., 2016), presumably because they disengage from the actual driving task. This translates also to other domains, as e.g. consultants struggled to maintain quality in their reports when they were initially written by an LLM. Compared to when consultants do the writing themselves, one study observed a 23% drop in the quality of problem-solving when consultants had the LLM write the first draft of the report (Candelon et al., 2023).

While we could of course try to remedy these situations by introducing a range of warning signals, a better solution would be to redesign the socio-technical system as a whole by giving the AI system a different role. When driver support systems in cars are designed not to take over the driving entirely, but rather give haptic feedback (changing the pressure in the gas pedal and nudging the steering wheel) in line with what the AI calculates as the optimal speed and direction, no reduction in response times is observed. In the very different setting of breast cancer screening Dembrower et al. (Dembrower et al., 2023) found that giving AI a role completely independent from the human operator can also be beneficial. Instead of the standard of two radiologists doing the screening, they tested a setup where the screening is done by one radiologist and (independently) by one AI system. Whenever one of the two flags potential breast cancer, the patient was then referred to the hospital for further testing, leading to both a big efficiency gain and an improvement in screening sensitivity.

The red thread we see in these examples is that in the first two, where we see performance drops, people have as their main role the correction of an AI system. Since failures are hard to detect, and it is additionally difficult to maintain attention when failures are rare, this leads to problems: opportunities to exercise our own skills are less frequent, and we thus become slowly more dependent on the AI systems to carry out the tasks for us. On the other hand, in the case of haptic control and the radiology screening we have a situation where humans carry out the entirety of the task, supported (in different degrees and ways) by an AI system. The person in this driving case actively has to turn the wheel and regulate speed, although aided by haptic cues from the automation. The radiologist has no direct interaction with the AI at all, thus keeping up their skills in the same way as without automation. Yet through the introduction of AI the socio-technical system as a whole is still improved. A different way to put this is that in the first cases we see a joint task scenario, where humans and AI systems solve the same task, with humans taking the role of correcting the AI at that task. Instead, we recommend to focus on designing systems where humans have independent tasks, not correcting the AI system but rather focusing on complementary work that together with the AI adds up to a more effective socio-technical system. In a practical implementation of these ideas, Carmona-Díaz et al. (Carmona-Díaz et al., 2025) present a tutorial for unstructured text analysis where LLMs are used for time-intensive classification tasks while human researchers maintain control over evaluation at multiple steps of the iterative process. Researchers can choose which stages of the process to automate and which to leave in humans' hands, to ensure that human expertise guides the process.

As a final note on this design point, we want to stress that this also means that we should design the socio-technical system in a way that prevents challenges to autonomy from outside as well. The NarxCare system mentioned above is a good example of how this can go wrong. In this case, doctors who prescribe more than the AI system advises are labeled as overprescribers. These doctors are then at risk of losing their medical license and may even face legal prosecution (Pozzi, 2023), thus creating a very high barrier for doctors to act differently than the system suggests. This harm to autonomy is not one stemming directly from the interaction between the AI and the doctor: it is one that is baked into the socio-technical system and points to a certain role that is ascribed to the AI system (as presumably being more objective and accurate, so that doctors who 'overprescribe' are seen as suspicious). It relegates users of the system to accepting the (negative) outputs of the system, thus taking away their autonomy. Designing the interaction in ways that disempower the human and place AI as an "automatic authority" (Lazar, 2024) are bound to amplify the negative deskilling effects discussed, as well as increase automated errors, rendering the sociotechnical system more fragile overall.

### 3.3. Protect Competency Skills by Designing Training Regimens

The redesign of socio-technical systems detailed above can help to protect both synchronic and diachronic competence. By ensuring that people do independent tasks, they maintain the skills to do those tasks and have the requisite environment to acquire them. This may not always be an option, as e.g. LLMs can only aid effectively in coding tasks when they are working in a joint task setting, working on the same code the human is working on. Of course, we might relegate an LLM to debugging, but in practice it is often useful to let it draft new code as well as debug code. In these kinds of settings, training regimens are a practical way to ensure that skills are maintained. These are common in high-stakes professions, as e.g. doctors, financial controllers and pilots are required to maintain and update their skills every year through such regimens. We may thus maintain competencies in a similar manner as we introduce AI into workflows, asking the humans involved to solve the task independently with some regularity.

Whether this is necessary will depend on a few factors. First, there is of course the question whether the socio-technical system as it is set up will indeed lead to a loss of competence through deskilling. Even if it doesn't, there is a question whether training regimens can help mitigate a loss of competence due to a lack of failure transparency. To our knowledge, empirical results on this point are not yet available, and so it is an open question whether such AI-specific training will help.

Second, there is a question regarding the type of task that we might become less competent in. The use of calculators has made us less competent in mental arithmetic, and this is rarely a problem. For once we acquire the necessary facility with numbers from doing arithmetic it seems fine that our mental arithmetic skill declines through long-term use of calculators. There is, however, a disanalogy with AI because calculators are both far more reliable and (serious) failures are easier to spot. As a result, we don't require training in calculations without calculators even for high-risk tasks, but there is a need for such training in high-risk joint tasks with AI.

Third, the need for such training regimens is likely to be higher when we focus on skill acquisition. As discussed, this can be hindered by AI support. Unsurprisingly, we don't learn to program nearly as well when an LLM produces first drafts of the code, just as we don't use calculators while learning mental arithmetic. There is, thus, a need to identify which skills remain crucial even as we may relegate some of them to AI systems later on. Mental arithmetic is still relevant because it leads to general numeracy skills. Coding will still be relevant even if more and more code is computer-generated, because it leads to insights into how computer programs function that are needed to determine code effectiveness, diagnose problems and brainstorm for new functionalities. Summing up, then, introducing additional training regimens can be crucial to ensure that we acquire and maintain essential skills. The human-AI interactions discussed here are far closer to the pilot-automation interactions in planes, that require extensive training and upkeep, than our use of calculators. Good design of training regimens is thus an important step in maintaining autonomy and preventing a diachronic loss of competency.

### 3.4. Protect Authenticity by Designing Adaptive Systems with Positive Friction

Finally, we argued that failures of authenticity can arise from *unconscious* AI-mediated value shifts of a decision-maker that *persist* after the human-AI interaction concludes. In particular, those from non-Western (Non-WEIRD) backgrounds and marginalized communities are particularly at risk of having their authenticity undermined because of the largely WEIRD values, beliefs, and biases embedded in systems, such as LLMs. To tackle these challenges, we propose that systems be designed with a level of *positive friction* to encourage cultivation of self-knowledge and designed to be *sufficiently adaptive* to a plurality of different value sets.

Regarding the former, positive friction could be used in AI to promote self-knowledge as a means of safeguarding oneself from unconscious value shifts. While there are varying accounts of self-knowledge, for our purposes we are using the term in the broadest sense to mean awareness of one's emotions, thoughts, beliefs, desires, and values (Gertler, 2010). In this regard, promoting *reflection* in technological design could be a helpful means of increasing self-knowledge. This could be achieved through the selective use of friction in the design of AI interactions. Researchers in the UX and HCI communities have begun exploring the role of design friction - design choices that inhibit the "flow" and ease at which a user navigates an interface - to promote more conscious, and autonomous interactions with technology (Cox et al., 2016). They utilize insights from behavioral and cognitive psychology to guide their exploration of positive friction - for example, the fast and slow processing model (Kahneman, 2011). In particular, in the field of data privacy decision-making, such design friction has been proposed as a means to help decision-makers make more reflective, value-centered privacy decisions in a cognitively overwhelming environment (Carter, 2022; Terpstra et al., 2019). Similarly, in AI, the use of such friction has also been proposed as a means of supporting one's self-control and goal-pursuit abilities, serving here as a reflective tool (Chen & Schmidt, 2024).

While more empirical work is needed to fully explore the efficacy of these interventions at supporting reflection, positive design friction is a promising means of cultivating greater self-knowledge by bringing one's unconscious interaction with technology to the forefront. In theory, bringing technology interactions to the decision-maker's awareness could have a protective effect on unconscious value shifts. By making the unconscious more known, one should be more able to deliberately endorse value shifts from the perspective of one's own practical identity, effectively safeguarding (some of) one's authenticity.

Next, we turn to ensuring that systems are sufficiently adaptive to promote authentic identity formation - that is, value shifts that are endorsed by the agent's core identity. Besides promoting conscious awareness of value shifts through positive friction, the systems themselves could be adaptive to the values of each collaborator as a further means of guarding against identity-inconsistent value shifts. While the positive friction would, ideally, promote a degree of self-knowledge that would allow someone interacting with an AI to consciously endorse value shifts, a system that too greatly varies from the agent's own value set could introduce a cognitive load that is simply too excessive for one to consciously endorse each and every shift. A sufficiently adaptive system, which can shift itself instead of shifting its human collaborator, could work in conjunction with friction to safeguard authenticity.

To create such a system will require understanding which values are currently instilled in AI (which can often be done through correlational studies, such as (Verma et al., 2025), without the need for new explainable AI techniques)[3]. As we discussed earlier, the values in many leading AI systems are largely WEIRD. In order to have a more active say in the values being instilled in AI, how values can - and which ones *should* - be incorporated into these systems has been an active area of research in the fields of value-alignment and value-sensitive design. In the value alignment arena, Anthropic's Claude takes a constitutional AI approach that seeks to align with broadly shared public values such as care and respect (Bai et al., 2022). Others have proposed using a human-rights based approach to align models (Prabhakaran et al., 2022) or identifying broad, universal values for aligning recommender systems (Stray, 2020). Similarly, the field of value-sensitive design aims to conceptualize and operationalize consensual norms and values in the design process through a tripartite methodology that includes stakeholder value conceptualization, empirical investigations, and technical investigations (Friedman et al., 2013).

However, in order for a system to be sufficiently adaptive and work to safeguard authenticity, it will need to be able to incorporate variation of values based on one's culture and lived experience. There are a number of current efforts to elicit and capture value plurality in AI which could help us accomplish such a feat, such as the concept of situated values (Arzberger et al., 2024) or personal values (Carter, 2024). Of course, those whose authenticity is most at risk from identity-inconsistent value shifts are those whose values are not currently being captured by AI systems. More work will therefore also be required to capture traditionally marginalized perspectives and values around AI, and designing adaptive systems to these value sets should be prioritized.[4]

## *4. Conclusion*

There is a lot to gain from human-AI interactions. Yet, at the same time, these potential benefits come at a risk to our domain-specific autonomy. As we have highlighted, through the lens of autonomy as self-governance, we see risks to both domain-specific competence and authenticity. Concerns around competence arise due to a lack of failure transparency and deskilling within domains, and around authenticity due to changes in our values resulting from interactions with AI systems. Each of these risks pose a challenge for human-AI interactions that need to support us not just in the short term, but also in long-term interactions. Especially the diachronic risks, of deskilling through loss of skills or ineffective skills acquisition and of inauthentic value changes, require much more attention to ensure that we maintain our autonomy to accomplish specific goals as we integrate AI into more and more aspects of our lives.

[3] Our thanks to a reviewer for pointing this out

[4] Notably, one should proceed carefully when adapting such systems to values in a personalized manner. As explored at great depth in Kirk et. al. (2024), the benefits of such personalization for autonomy and, as we have argued, specifically *authenticity,* will need to be carefully balanced against other ethical concerns around data privacy and algorithmic bias.

At the same time, we believe that good design of human-AI interactions can help to mitigate these risks. The role that AI systems are assigned in human-AI interactions matter greatly, and we've highlighted that in particular joint task settings, where humans and AI are effectively solving the same task, are likely to lead to reduced human autonomy for those tasks. Instead, independent task setups where humans and AI systems solve complementary tasks already avoid many of the issues we have highlighted. That being said, we still need further steps to protect both our competencies and our authenticity. We can make failures of AI systems more transparent, for example by actively designing for defeaters (Buijsman & Veluwenkamp, 2022). We can also ensure that training regimens are in place that maintain skills and ensure effective acquisition of new skills. Here, temporary suspension of the AI system may lead to benefits in the long run.

Finally, we need to actively ensure that our values remain authentic. Positive friction in human-AI interactions can promote reflection on our values and ensure that any shifts in those values happen consciously and authentically. More adaptive systems that better align to the personal values of the human it is interacting with can also help to prevent the scenario where humans adapt their values to suit those of a more rigid AI system. Both these steps can help prevent inauthentic value shifts that can often be seen as a type of manipulation that goes beyond individual actions and deserves more attention. Only if we tackle all these parts of autonomy in the design of human-AI interactions can we ensure that we benefit from them in the long run.